\documentclass[aps, prd, twocolumn, lengthcheck, superscriptaddress, 
nofootinbib]{revtex4-1}

\usepackage{color}
\usepackage{hyperref}
\usepackage{ulem}
\usepackage{graphicx}
\usepackage{mathrsfs}
\usepackage{amsmath}

\def\be{\begin{eqnarray}}
\def\ee{\end{eqnarray}}

\newcommand*{\nm}{\nonumber}

\begin{document}
\title{Two-flavor color superconductivity in a general Nambu-Jona–Lasinio model with color and charge neutrality}

\author{Li-Kang Yang}

\author{Di-Sheng Fan}
\affiliation{School of Physics, Nanjing University, Nanjing, 210093, China}

\author{Cheng-Ming Li}
\affiliation{Institute for Astrophysics, School of Physics, Zhengzhou University, Zhengzhou 450001, China}

\author{Yong-Liang Ma}
\email{ylma@nju.edu.cn}
\affiliation{School of Frontier Sciences, Nanjing University, Suzhou, 215163, China}
\affiliation{International Center for Theoretical Physics Asia-Pacific (ICTP-AP) , UCAS, Beijing, 100190, China}



\begin{abstract}
By using a general NJL model including as many interaction channels as possible, and taking into account the constraints imposed by color and charge neutrality, we analyze how the different interaction channels contribute to the charge-neutral two-flavor color-superconducting matter. After taking the Fierz transformation, the number of parameters in the NJL model is reduced and thereby quark-antiquark condensates and diquark condensates are related. Through a self-consistent solution of the gap equations, we find that in addition to the diquark and vector channels, the scalar-isovector, vector-isovector, and vector-isovector-color-octet channels are also important, while other channels can be neglected. Moreover, between the normal quark matter phase and two-flavor color-superconducting phase, there is a gapless two-flavor color-superconducting phase. The fractions of quarks with different flavors and colors are calculated in both gaped and gapless two-flavor color-superconducting phases. In addition, We illustrate the phase diagrams in the diquark coupling and chemical potential plane, and discuss in detail how the dominant channels influence these phase diagrams. 
\end{abstract}

\maketitle
\allowdisplaybreaks{}


\section{Introduction}
The notion of color superconductivity (CSC) was introduced in the 1970s~\cite{PhysRevLett.34.1353,BARROIS1977390,Frautschi1980}, in analogy to the Bardeen-Cooper-Schrieffer (BCS) theory in condensed matter physics~\cite{PhysRev.104.1189,PhysRev.106.162,PhysRev.108.1175}. In BCS theory, phonon exchange provides a weak attractive force between electrons, while in CSC, quarks experience an attractive interaction mediated by gluon exchange in the color antitriplet channel. This attraction enables quarks near the Fermi surface to form Cooper pairs. 

Early studies of CSC claimed that the superconducting gap is only at the order of MeV~\cite{BAILIN1984325}, making it difficult to exhibit significant phenomenological effects. Consequently, this phenomenon initially received little attention. However, this situation changed in the 1990s after the subsequent research found that in the moderate density regions, the gap of CSC could reach up to $\approx 100$~MeV~\cite{PhysRevLett.81.53,ALFORD1998247}, substantially enhancing its physical significance. The CSC at high densities was analyzed using a weak coupling approach in Refs.~\cite{Iwasaki:1994ij,Pisarski:1999av,Pisarski:1999bf}. Since then, research in this field has flourished, with extensive investigations focusing on CSC in the moderate-density regimes and its implications for both phase structure of quantum chromodynamics (QCD)~\cite{ALFORD1999443,Ratti:2004ra,PhysRevD.72.034004,PhysRevLett.93.132001,PhysRevD.111.014006} and astrophysical observations~\cite{Roupas:2020nua,PhysRevD.72.065020,KLAHN2007170,PhysRevLett.119.161104,PhysRevLett.132.262701,Christian:2025dhe}. For the deeper insights into the mechanisms and implications of CSC, we refer to some review articles, e.g.,~\cite{PhysRevD.63.074016,RevModPhys.80.1455,doi:10.1142/9789812810458_0043, RevModPhys.86.509}.

Color superconducting matter exhibits rich phase structures, a consequence of its diverse quark pairing patterns~\cite{RevModPhys.86.509,BUBALLA2005205,wang2010some,PhysRevD.71.054016}. Among these pairing patterns, the color-flavor locked (CFL) phase is regarded as the ground state of QCD matter at asymptotic density~\cite{ALFORD1999443,SCHAFER2000269,SHOVKOVY1999189}. In the moderate-density regime, due to the  substantial mass difference between strange quark and up/down quarks, when the chemical potential is below the order of the strange quark mass, the CFL phase may be disfavored and replaced by a two-flavor color superconducting phase (2SC)~\cite{ALFORD1999219,PhysRevLett.87.062001,doi:10.1142/S0218301305003491}. When studying the CSC in electrically neutral quark matter, such as the core of neutron stars, the phase diagram becomes even richer. The standard 2SC phase may evolve into a gapless 2SC (g2SC) phase with gapless quasiparticle excitations~\cite{HUANG2003835,SHOVKOVY2003205,Shovkovy:2004me}. In the transition region between the 2SC and CFL phases, especially at finite temperatures, gapless CFL (gCFL) and superconductivity including up quark (uSC) phases are expected to emerge~\cite{PhysRevD.72.034004,PhysRevLett.93.132001}. 

The g2SC phase has recently attracted significant attention due to its stability. Studies shown that the g2SC phase exhibits thermal stability under the constraint of local charge neutrality~\cite{HUANG2003835,SHOVKOVY2003205}, but it also suffers from chromomagnetic instability associated with the Meissner effect~\cite{PhysRevD.70.051501,PhysRevD.70.094030}. The chromomagnetic instability arises not only in the g2SC phase but also in the standard 2SC phase when the mismatch between the chemical potentials of up and down quarks, $|\delta\mu|$, exceeds $\Delta/\sqrt{2}$ with $\Delta$ being the 2SC gap (its explicit definition will be given later)~\cite{PhysRevD.70.051501}. This instability suggests that the g2SC phase may not be a stable state but could transition to other phases, such as inhomogeneous phases like the crystalline phase~\cite{RevModPhys.86.509,GIANNAKIS2005137,GIANNAKIS2005255,PhysRevLett.96.022005}, solitonic phase~\cite{PhysRevD.79.054009}, the gluonic phase~\cite{GORBAR2006305,PhysRevD.73.111502}, and so on. Consequently, further research is needed to clarify the exact nature and implications of these instabilities.

In the literature, the 2SC and g2SC phases have been studied within the framework of the Nambu-Jona–Lasinio (NJL) model~\cite{Nambu:1961tp,Nambu:1961fr}, a powerful model used in a variety of phenomena of strong interaction at low energy~\cite{Vogl:1991qt,Klevansky:1992qe,Hatsuda:1994pi,BUBALLA2005205}. 
The NJL model has many interaction channels, and their effects on the phase diagram and neutron star physics have been widely studied in, e.g., Refs.~\cite{PhysRevD.72.034004,PhysRevD.94.065032,PhysRevC.101.055204,KITAZAWA2003C289,PhysRevD.100.094012}. The influence of diquark channels on the CSC phase diagram was studied in Refs.~\cite{HUANG2003835,PhysRevD.72.034004}. And the effects of the diquark channel combining with the vector and vector-isovector channels on the equations of state (EOS) of the neutron stars were investigated in Refs.~\cite{KLAHN2007170,PhysRevC.101.055204,KITAZAWA2003C289}. The results show that in addition to the diquark channel, the vector and isovector channels also have significant effects on the EOS of neutron stars and the CSC phase diagram. In addition to the NJL models, the phase structure of CSC matter was also be studied by using the renormalization group approach or the Schwinger-Dyson equation~\cite{Arnold:1998cy,Son:1998uk,Evans:1998nf,Hong:1999fh}.

In this work, we systematically investigate the effects of all the possible channels, particularly those often overlooked in the literature, in the most general NJL model on the QCD phase diagram and the EOS for neutron stars. The general NJL model has many interaction channels, especially those could influence the charge-neutral 2SC phase. These channels are not independent but can be connected through a Fierz transformation. We further use the one-gluon-exchange (OGE) model to obtain the coupling constants of each channel. By imposing the electro-charge and color-charge neutrality which were not seriously considered before~\cite{PhysRevD.65.014018,PhysRevD.108.043008}, we analyze the contributions of each channel in more detail and identify the dominant interaction channels. Then we discuss their individual impacts on the phase diagram beyond the OGE model by varying their couplings.

We found that, apart from the diquark and vector channels which have been extensively discussed in the literature, the scalar-isovector, vector-isovector, and vector-isovector-color-octet channels also play significant roles, whereas the contributions from other channels are negligible. This arises from enforcing color and charge neutrality constraints, which enhance the importance of these channels in determining the phase structure. In addition, due to the color and charge neutrality conditions, we find that in the OGE model the system evolves from the chiral breaking phase to the g2SC phase, and then to the 2SC phase. Compared to the case without the charge-neutrality, including charge neutrality introduces an additional intermediate g2SC phase, making the phase diagram more complex. Moreover, we examined how the ratio of the diquark-channel coupling constant $H$ to the scalar-channel coupling constant $G_s^{(0)}$ influences the 2SC phase structure. We estimated the critical endpoint (CEP) at $H/G_s^{(0)} = 1.1$, $\mu = 312\,\text{MeV}$. 
The g2SC phase is delineated by two phase boundaries, corresponding to  $H/G_s^{(0)} \approx 0.67$ and $H/G_s^{(0)} \approx 0.81$.
For the effects of distinct channels, we found that the vector channels shift the entire phase diagram toward larger chemical potential ($\mu$), while the scalar-isovector channel extends the gapless phase region from $0.67 \lesssim H/G_s^{(0)} \lesssim 0.80$ to $0.67 \lesssim H/G_s^{(0)} \lesssim 0.85$ in the vicinity of the chiral phase transition.
Moreover, the isovector channel lowers the values of $H/G_s^{(0)}$ corresponding to both the upper and lower boundaries of the g2SC phase.

The rest of this paper is organized as follows. In Sec.~\ref{sec:NJL}, we derive the general NJL model and give the coupling relations by Fierz transformation. In Sec.~\ref{sec:DC}, we figure out the dominant channels in the OGE model. In Sec.~\ref{sec3}, we discuss the effects of these channels on the 2SC phase structure separately at zero temperature. Finally, we make conclusions in Sec.~\ref{sec4}.

\section{General framework}
\label{sec:NJL}

The most general Lorentz invariant Lagrangian of the four-fermion interaction NJL model
can be decomposed as~\cite{KLIMT1990429}
\be
\mathcal{L}^{(4)}_{\rm NJL} & = & \mathcal{L}_{\rm sym}+\mathcal{L}_{\rm det}, 
\label{eq:GeneralNJL}
\ee
where $\mathcal{L}_{\rm sym}$ is the Lagrangian preserving $SU(N_c)\times SU(N_f)_V\times SU(N_f)_A\times U(1)_V\times \{C,P,T\}$ symmetry and $\mathcal{L}_{\rm det}$ breaks $U(1)_A$ symmetry. They can be generally expressed as
\begin{widetext}
\begin{subequations}
\be
\mathcal{L}_{\rm sym} & = & \bar{\psi}_a\psi_b\bar{\psi}_c\psi_d\left[C_1(SS+PP)\otimes(t_0t_0+tt)\otimes\delta^C\delta^C+C_2(VV+AA)\otimes(t_0t_0+tt)\otimes\delta^C\delta^C\right. \nm\\
& & \left.\qquad\qquad\;\; {} +C_3(VV+AA)\otimes t_0t_0\otimes\delta^C\delta^C+C_4(VV-AA)\otimes t_0t_0\otimes\delta^C\delta^C\right]_{ab,cd} \nm\\
& &{} + \bar{\psi}_a\psi_b\bar{\psi}_c\psi_d\left[C_5(SS+PP)\otimes(t_0t_0+tt)\otimes\lambda^C\lambda^C+C_6(VV+AA)\otimes(t_0t_0+tt)\otimes\lambda^C\lambda^C\right.\nm\\
& &\left.\qquad\qquad\qquad {} + C_7(VV+AA)\otimes t_0t_0\otimes\lambda^C\lambda^C+C_8(VV-AA)\otimes t_0t_0\otimes\lambda^C\lambda^C\right]_{ab,cd}, \label{eq:NJLsym}\\
\mathcal{L}_{\rm det} & = &{} G\left(\text{det}\left[\bar{\psi}(1+\gamma^5)\psi\right]+\text{det}\left[\bar{\psi}(1-\gamma^5)\psi\right]\right),
\label{eq:NJLdet}
\ee
\end{subequations}
where
$\{S,P,V,A\}=\{1,i\gamma_5,\gamma^\mu,\gamma^\mu\gamma_5\}$ are the Dirac matrices, $\{t_0,t_i\}$ are generators $U(N_f)$ with $\text{Tr}\left[t_at_b\right]=2\delta_{ab}$ and $\{\delta^C,\lambda^C\}$ are the identity and Gell-Mann matrices in color space. 

Some of the interaction terms in Lagrangian~\eqref{eq:GeneralNJL} are not independent but relate to each other by Fierz transformation through exchanging $\psi_a$ and $\psi_c$ or $\psi_b$ and $\psi_d$. The interactions related by Fierz transformation 
are in fact dynamically equivalent. Therefore, independent interaction terms form a Fierz complete basis, i.e., the minimal set of interactions which can generate 
all other interactions by Fierz transformation~\cite{PhysRevD.96.076003,KLIMT1990429}.
Thus, we choose a Fierz complete interaction set
\be
\mathcal{L}_{\rm sym}^{\rm FCI} & = & \bar{\psi}_a\psi_b\bar{\psi}_c\psi_d\left[C_1^\prime(SS+PP)\otimes(t_0t_0+tt)\otimes\delta^C\delta^C+C_2^\prime(VV+AA)\otimes(t_0t_0+tt)\otimes\delta^C\delta^C\right.\nm\\
& &\left.\qquad\qquad\;\; {} +C_3^\prime(VV+AA)\otimes t_0t_0\otimes\delta^C\delta^C+C_4^\prime(VV-AA)\otimes t_0t_0\otimes\delta^C\delta^C\right]_{ab,cd}.
\label{eq:FirezBasis}
\ee
Then, after Fierz transformation, coupling constants $C_i^\prime$ can be related to $C_i$. We can also rewrite the Lagrangian~\eqref{eq:NJLsym}
into a Fierz invariant form as follows
\be
\mathcal{L}^{\mathscr{F}}_{\rm sym} & = & \frac{1}{2}\left[\mathcal{L}_{\rm sym}^{\rm FCI}+\mathscr{F}_{\bar{q}q}(\mathcal{L}_{\rm sym}^{\rm FCI})\right] \nm\\
& = & \frac{1}{2}\bar{\psi}_a\psi_b\bar{\psi}_c\psi_d\left[G_1(SS+PP)\otimes(t_0t_0+tt)\otimes\delta^C\delta^C+G_2(VV+AA)\otimes(t_0t_0+tt)\otimes\delta^C\delta^C\right. \nm\\
& & \left.\qquad\qquad\quad\; {} +G_3(VV+AA)\otimes t_0t_0\otimes\delta^C\delta^C+G_4(VV-AA)\otimes t_0t_0\otimes\delta^C\delta^C\right]_{ab,cd} \nm\\
& &{} + \frac{N_c^2}{N_c^2-1}\bar{\psi}_a\psi_b\bar{\psi}_c\psi_d\left[\left(-\frac{1}{2N_c}G_1-\frac{1}{N_f}G_4\right)(SS+PP)\otimes(t_0t_0+tt)\otimes\lambda^C\lambda^C\right. \nm\\
& &\qquad\qquad\qquad\qquad\qquad{} +\left(-\frac{1}{2N_c}G_2+\frac{1}{2N_f}G_3\right)(VV+AA)\otimes(t_0t_0+tt)\otimes\lambda^C\lambda^C \nm\\
& &\qquad\qquad\qquad\qquad\qquad{} +\left(\frac{N_f}{2}G_2-\frac{1}{2N_c}G_3\right)(VV+AA)\otimes t_0t_0\otimes\lambda^C\lambda^C \nm\\
& & \left.\qquad\qquad\qquad\qquad\qquad{} +\left(-\frac{N_f}{4}G_1-\frac{1}{2N_c}G_4\right)(VV-AA)\otimes t_0t_0\otimes\lambda^C\lambda^C\right]_{ab,cd},
\ee
\end{widetext}
where $\mathscr{F}_{\bar{q}q}$ denotes the quark-antiquark channel Fierz transformation and the $G_1\sim G_4$ relate to $C_1^\prime\sim C_4^\prime$ through
\be
G_1 & = & C_1^\prime-\frac{2}{N_fN_c}C_4^\prime ,\quad G_2=C_2^\prime+\frac{1}{N_fN_c}C_3^\prime,\nm\\
G_3 & = & C_3^\prime + \frac{N_f}{N_c}C_2 ,\quad G_4=C_4^\prime-\frac{N_f}{2N_c}C_1^\prime.
\ee
Similarly, we can obtain the relations between $G_i$ and $C_i$. In this sense, the eight parameters in~\eqref{eq:NJLsym} are reduced to four.

Note that any two NJL-type four-fermion interactions that are both chiral and $U(1)_A$ invariant should be physically equivalent if they predict the same values of $G_1\sim G_4$ in the Fierz invariant form. Then, combining the determinant interaction, 
the general NJL model only has five independent couplings $G_1\sim G_4$ and $G$.
For example, the original NJL model~\cite{Nambu:1961tp} can be obgained by setting $C_1^\prime=G/2$ and $C_2^\prime=C_3^\prime=C_4^\prime=0$ in the Lagrangian~\eqref{eq:FirezBasis}. In the OGE model which are widely used in the study of CSC~\cite{ALFORD1999443,ALFORD1999219,PhysRevD.65.014018,PhysRevD.108.043008}
\be
\mathcal{L}_{\rm int}=-g(\bar{\psi}\gamma^\mu\lambda_a\psi)^2,
\label{eq:Lag1ge}
\ee
we have $G_1=-2G_2=(N_c^2-1)g/N^2_c$ and $G_3=G_4=G=0$.

In order to study the color superconducting phase structure, we should consider the
diquark interactions which can be derived via the Fierz transformation by the exchange of $\psi_b$ and $\psi_c$ or $\psi_a$ and $\psi_d$ based on the quark-antiquark channel interactions given in Eq.~(\ref{eq:FirezBasis}). Explicitly, the Lagrangian takes the form 
\begin{widetext}
\be
\mathscr{F}_{qq}\left[\mathcal{L}_{\rm sym}^{\rm FCI}\right] & = & \bar{\psi}_a\bar{\psi}_b\psi_c\psi_d\biggl\{H_1\left[(i\gamma_5C)(i\gamma_5C)+(C)(C)\right]\otimes t_At_A\otimes \lambda_A\lambda_A \nm\\
& & \qquad\qquad\quad {} + H_2\left[(i\gamma_5C)(i\gamma_5C)+(C)(C)\right]\otimes t_St_S\otimes \lambda_S\lambda_S \nm\\
& & \qquad\qquad\quad {} + H_3\left[(\gamma^\mu\gamma_5 C)(C\gamma_\mu\gamma_5)\otimes t_At_A +(\gamma^\mu C)(C \gamma_\mu)\otimes t_St_S\right]\otimes\lambda_A\lambda_A \nm\\
& & \qquad\qquad\quad {} +H_4\left[(\gamma^\mu\gamma_5 C)(C\gamma_\mu\gamma_5)\otimes t_St_S +(\gamma^\mu C)(C \gamma_\mu)\otimes t_At_A\right]\otimes\lambda_S\lambda_S
    \biggr\}_{ab,cd},
\ee
\end{widetext}
where $\mathscr{F}_{qq}$ denotes the diquark channel Fierz transformation, the indices $S$ and $A$ represent the symmetric and antisymmetric part of corresponding group generators and $C=i\gamma^2\gamma^0$. The couplings $H_i$ relate to $G_i$ through
\be
H_1 & = & \frac{N_c}{1-N_c}\left[G_2-\frac{1}{N_f}G_3\right], \nm\\
H_2
&  = &\frac{N_c}{1+N_c}\left[G_2+\frac{1}{N_f}G_3\right], \nm\\
H_3
& = & \frac{N_c}{1-N_c}\left[\frac{1}{4}G_1+\frac{1}{2N_f}G_4\right], \nm\\
H_4
& = & \frac{N_c}{1+N_c}\left[\frac{1}{4}G_1-\frac{1}{2N_f}G_4\right].
\ee
Note that $\mathcal{L}_{\rm sym}^{\rm FCI}$, $\mathscr{F}_{\bar{q}q}\left[\mathcal{L}_{\rm sym}^{\rm FCI}\right]$ and $\mathscr{F}_{qq}\left[\mathcal{L}_{\rm sym}^{\rm FCI}\right]$ are, in fact, equivalent. We only extract their distinct interaction channels through the Fierz transformation.

Although parameters $G_1\sim G_4$ can be partly determined by the masses and decay constants of mesons in vacuum~\cite{KLIMT1990429}, in this work, in order to 
investigate the influence of different channels on the 2SC phase structure, we vary their values except the coupling $G^{(0)}_s$ of the scalar interaction $(\bar{\psi}\psi)^2$ which is fixed by the chiral symmetry breaking and the pion properties in vacuum. 

Due to the Fierz transformation, the Hartree-Fock and BCS mean field approximation (MFA) of model $\mathcal{L}_{\rm sym}+\mathcal{L}_{\rm det}$ can be written as
\begin{widetext}
\be
\langle\mathcal{L}_{\rm sym}+\mathcal{L}_{\rm det}\rangle_{\rm HFB} & = & \langle\mathcal{L}_{\rm sym}+\mathcal{L}_{\rm det}\mathcal\rangle_{\rm H} + \langle\mathscr{F}_{\bar{q}q}\left[\mathcal{L}_{\rm sym}+\mathcal{L}_{\rm det}\right]\mathcal\rangle_{\rm H} + \langle\mathscr{F}_{qq}\left[\mathcal{L}_{\rm sym}+\mathcal{L}_{\rm det}\right]\rangle_{\rm B},
\label{eq:HFB}
\ee
where the sub-indices ${\rm HFB}, {\rm H}$ and ${\rm B}$ represent the Hartree-Fock and BCS MFAs, Hartree MFA and BCS MFA, respectively.

Although the NJL model includes many interaction channels, after MFA, only few of them contribute due to the symmetry consideration. In the charge neutral 2SC phase, the following channels survive
\be
\label{eq8}
\mathcal{L}_{\bar{q}q} & = & G^{(0)}_s\left(\bar{\psi}\psi\right)^2+G^{(8)}_s\left(\bar{\psi}\lambda_a \psi\right)^2+G^{(0)}_{sIv}\left(\bar{\psi}\tau \psi\right)^2 + G^{(8)}_{sIv}\left(\bar{\psi}\tau \lambda_a \psi\right)^2+G^{(0)}_v\left(\bar{\psi}\gamma^0 \psi\right)^2 \nm\\
& &{} + G^{(8)}_v\left(\bar{\psi}\gamma^0\lambda_a \psi\right)^2+G^{(0)}_{vIv}\left(\bar{\psi}\gamma^0\tau_3\psi\right)^2 + G^{(8)}_{vIv}\left(\bar{\psi}\gamma^0\tau_3\lambda_a\psi\right)^2,\nm\\
\mathcal{L}_{qq} & = & H\left(\bar{\psi}i\gamma_5\tau_2\lambda_2\psi^C\right)\left(\bar{\psi}^Ci\gamma_5\tau_2\lambda_2 \psi\right) + H_0\left(\bar{\psi}\gamma^0\gamma_5\tau_2\lambda_2\psi^C\right)\left(\bar{\psi}^C\gamma^0\gamma_5\tau_2\lambda_2 \psi\right),
\ee
where the upper indices $(0)$ and $(8)$ denote the color singlet and octet, respectively, and the lower indices $s, sIv, v$ and $vIv$ represent the Lorentz-scalar, Lorentz-scalar isovector, Lorentz-vector and Lorentz-vector-isovector. Here, we did not consider the color-sextet diquark channels since they are in the the repulsive channels and the condensates are small~\cite{ALFORD1999443,SHOVKOVY1999189}. In addition, in 2SC, we preserve the color symmetry between red and green quarks. The couplings in~\eqref{eq8} relate to $G$ and $G_i$ through
\be
\label{eq9}
G^{(0)}_s & = & G_1+\frac{1}{2}G+\frac{1}{4N_c}G, \nm\\
G^{(8)}_s & = & \frac{N^2_c}{N^2_c-1}\left(-\frac{1}{2N_c}G_1-\frac{1}{N_f}G_4\right)+\frac{G}{8}, \nm\\
G^{(0)}_{sIv} &  = & G_1-\frac{1}{2}G-\frac{1}{4N_c}G, \nm\\
G^{(8)}_{sIv} & = & \frac{N^2_c}{N^2_c-1}\left(-\frac{1}{2N_c}G_1-\frac{1}{N_f}G_4\right)-\frac{G}{8}, \nm\\
G^{(0)}_v & = & G_2+G_3+G_4, \nm\\
G^{(8)}_v & = & \frac{N^2_c}{N^2_c-1} \left(-\frac{1}{2N_c}G_2+\frac{1}{2N_f}G_3+\frac{N_f}{2}G_2 -\frac{1}{2N_c}G_3-\frac{N_f}{4}G_1
-\frac{1}{2N_c}G_4\right), \nm\\
G^{(0)}_{vIv} & = & G_2, \nm\\
G^{(8)}_{vIv} & = & \frac{N^2_c}{N^2_c-1}\left(-\frac{1}{2N_c}G_2+\frac{1}{2N_f}G_3\right), \nm\\
H & = & H_1+\frac{G}{8}, \nm\\
H_0 & = & H_3.
\ee


\section{Contribution from different channel}
\label{sec:DC}

In this section, we investigate the effect of individual interactions in Eq.~(\ref{eq8}) on the color superconducting phase. 

\subsection{Thermal potential}

From Eq.~\eqref{eq:HFB}, we obtain the following thermodynamic potential
\begin{equation}
\Omega=\Omega_M+U+\Omega_e,
\end{equation}
where $\Omega_M$, $U$ and $\Omega_e$ are the contributions from the quark dynamical part, the quark condensate part, and the electron part, respectively. They are expressed as
\be
\label{eq11}
\Omega_M & = &{} -\sum_{i=1}^4\int^\Lambda_0\frac{p^2dp}{\pi^2}\left[\epsilon^i_{r}+2T\ln\left(1+e^{-\epsilon^i_{r}/T}\right)\right] \nm\\
& &{} -\sum_{f=u,d}\int^\Lambda_0\frac{p^2dp}{\pi^2}E_{fb}+T\ln\left[1+e^{-\left(E_{fb}+\hat{\mu}_{fb}\right)/T}\right]+T\ln\left[1+e^{-\left(E_{fb}-\hat{\mu}_{fb}\right)/T}\right] ,\nm\\
U & = & G^{(0)}_s\phi^2+G^{(8)}_s\phi^2_8+G^{(0)}_{sIv}\phi_I^2+G^{(8)}_{sIv}\phi_{I8}^2+G^{(0)}_vn^2+G^{(8)}_v n^2_8+G^{(0)}_{vIv}n_I^2+G^{(8)}_{vIv}n_{I8}^2 \nm\\
& &{} + H\left|\delta\right|^2+H_0\left|\delta_0\right|^2 ,\nm\\
\Omega_e & = &{} -\int\frac{p^2 dp}{\pi^2}\left[T\ln\left(1+e^{-(p+\mu_Q)/T}\right)+T\ln\left(1+e^{-\left(p-\mu_Q\right)/T}\right)\right].
\ee
Here, $E_{fc}=\sqrt{p^2+M^2_{fc}}$ for $f=\{u,d\},\,c=\{r,g,b\}$. $\phi=\langle \bar{\psi}\psi\rangle, \phi_8=\langle \bar{\psi}\lambda_a\psi\rangle, \phi_I=\langle \bar{\psi}\tau_3\psi\rangle$ and $\phi_{I8}=\langle \bar{\psi}\tau_3\lambda_a\psi\rangle$ represent the Lorentz-scalar density in color-singlet, color-octet, iso-vector and isovector-color-octet channels, respectively. $n=\langle \bar{\psi}\gamma^0\psi\rangle, n_8=\langle \bar{\psi}\gamma^0\lambda_a\psi\rangle, n_I=\langle \bar{\psi}\gamma^0\tau_3\psi\rangle$ and $n_{I8}$ denote quark number density in color-singlet, color-octet, iso-vector and isovector-color-octet channels, respectively. $\delta=\langle\bar{\psi}^C \gamma_5\tau_2\lambda_2\psi\rangle$ is diquark condensate Lorentz-scalar channel and $\delta_0=\langle\bar{\psi}^C \gamma^0\gamma_5\tau_2\lambda_2\psi\rangle$ stands for the diquark condensate in Lorentz-vector channel. $\hat{\mu}_{fc}$ is the effective chemical potential for $f$-flavor $c-$color quark and $\mu_e={}-\mu_Q$ is the chemical potential of electrons. The spectra $\epsilon^i_{r}$ in red and green subspace is the eigenvalues of the characteristic equation 
\be
\label{eq12}
& & \left[(-\lambda+\delta \mu)^2 -p^2-\bar{M}_r^2-\bar{\mu}^2-\Delta^2 - \Delta_0^2 + (\delta M_r)^2 \right]^2 \nm\\
& &{} - 
    4 \left[ \left( -\delta M_r(-\lambda+\delta \bar{\mu}) + \bar{M}_r \bar{\mu}- \Delta\Delta_0\right)^2 + p^2 (\Delta_0^2 + \bar{\mu}^2- \delta M_r^2 )\right]=0,
\ee
\end{widetext}
where $\Delta=-2H\delta$, $\Delta_0=2H_0\delta_0$ and
\be
\bar{\mu} & = & \frac{\hat{\mu}_{ur}+\hat{\mu}_{dr}}{2},\;\; \delta \mu = \frac{\hat{\mu}_{ur}-\hat{\mu}_{dr}}{2},\nm\\
\bar{M}_r & = & \frac{M_{ur}+M_{dr}}{2},\;\; \delta M_r = \frac{M_{ur}-M_{dr}}{2}.
\ee
Due to the degeneracy between the red and green quarks, we have $M_{fg}=M_{fr}$ and $\hat{\mu}_{fg}=\hat{\mu}_{fr}$ in the thermodynamic potential~(\ref{eq11}) and in
 the characteristic equation~(\ref{eq12}). Thus, we obtain the following relations:
\be
\frac{\delta\Omega_M}{\delta M_{fr}} & = & \phi_{fr}+\phi_{fg}=2\phi_{fr}, \nm\\
{} -\frac{\delta \Omega_M}{\delta \hat{\mu}_{fr}} & = & n_{fr}+n_{fg}=2n_{fr}, 
\ee
with $f=u,\,d$ quarks.

In Refs.~\cite{SHOVKOVY2003205,HUANG2003835}, the criterion for a g2SC phase is $\Delta\leq|\delta\mu|$. But this criterion is not applicable here when we consider the contributions of $\delta M_r$ and $\Delta_0$. 
At present, the criterion for the g2SC phase is that one of the spectra $\epsilon^i_r(p)$---eigenvalues of the characteristic equation~\eqref{eq12}---becomes zero at some momentum $p$. This implies that Eq. (\ref{eq12}) has a real number solution about momentum $p$ when we take $\lambda=0$. Equivalently, 
\begin{widetext}
\be
\Delta_0^2 (\delta \mu^2 - \bar{M}_r^2) + (\bar{M}_r\delta M_r  - \bar{\mu}\delta \mu)^2 + 2 \Delta \Delta_0 (\delta M_r \delta \mu - \bar{M}_r \bar{\mu}) + 
 \Delta^2 (\delta M_r^2 - \bar{\mu}^2)\geq0.
 \label{eq:Crig2SC}
\ee
If both $\delta M_r$ and $\Delta_0$ vanish, the above inequality reduce to $\Delta\leq|\delta \mu|$.

When we take HFB MFA, the constituent quark mass $M_{fc}$, effective chemical potential $\hat{\mu}_{fc}$ and 2SC gaps $\Delta$, $\Delta_0$
are related to condensates and densities through
\be
\label{eq19}
{M}_{fc} & = & m-2G^{(0)}_s\phi-2G^{(8)}_s(\lambda_8)_{cc}\phi_8 -2G^{(0)}_{sIv}(\tau_3)_{ff}\phi_I - 2G^{(8)}_{sIv}(\tau_3)_{ff}(\lambda_8)_{cc}\phi_{I8},\nm\\
\hat{\mu}_{fc} & = & \mu+2G^{(0)}_{v}n+Q_{f}\mu_Q + (\lambda_8)_{cc}(\mu_{8}+2G^{(8)}_vn_8) + 2G^{(0)}_{vIv}(\tau_3)_{ff}n_I + 2G^{(8)}_{vIv}(\tau_3)_{ff}(\lambda_8)_{cc}n_{I8},
\ee
\end{widetext}
with $Q_u=2/3,\, Q_d=-1/3$ and $\lambda_8=\frac{1}{\sqrt{3}}{\rm diag} (1,1,-2)$. 
The condensates and densities in above equations can be obtained by the minimal of the thermodynamic potential 
\be
& & \frac{\delta \Omega}{\delta \phi }=\frac{\delta \Omega}{\delta \phi_8 } =\frac{\delta \Omega}{\delta \phi_I }=\frac{\delta \Omega}{\delta \phi_{I8} } =\frac{\delta \Omega}{\delta n }\nm\\
& &\;\;\;\;\;\;\, =\frac{\delta \Omega}{\delta n_8 }=\frac{\delta\Omega}{\delta n_I}=\frac{\delta\Omega}{\delta n_{I8}}=\frac{\delta \Omega}{\delta \delta }=\frac{\delta \Omega}{\delta \delta_0 }=0.
\ee
Because of the charge and color neutrality, we have 
\be
\label{eq21}
n_Q & = &{} -\frac{\delta\Omega}{\delta\mu_Q}=0 ,\nm\\
n_8 & = &{} -\frac{\delta\Omega}{\delta\mu_8}=0.
\ee
These neutrality conditions which have not been seriously considered before yield relations among the chemical potentials $\{\mu,\,\mu_Q\,\mu_8\}$, such that there is only one independent chemical potential $\mu$.

\subsection{Gap equation}

To self-consistently analyze the contribution from a specific channel, we solve the gap equations~\eqref{eq19}-\eqref{eq21} in the OGE model~\eqref{eq:Lag1ge} which has only independent coupling $g$. The other couplings relate to $g$ through Eq.~(\ref{eq9}). The model parameters are cutoff $\Lambda=631 \text{~MeV}$ in Eq.~\eqref{eq11}, current quark mass $m=5.5\text{~MeV}$ in Eq.~\eqref{eq19} and $G^{(0)}_s=2.188/\Lambda^2$~\cite{MASAYUKI1989668}, which were obtained by fitting 
$\langle\bar{u}u\rangle^{1/3}=-246\text{~MeV}$, $f_\pi=93 \text{~MeV}$ and $m_\pi=138\text{~MeV}$. The coupling $g$ is chosen such that $G^{(0)}_s=2.188/\Lambda^2$.
\begin{figure}[htpb]
    \centering
\includegraphics[scale=0.5]{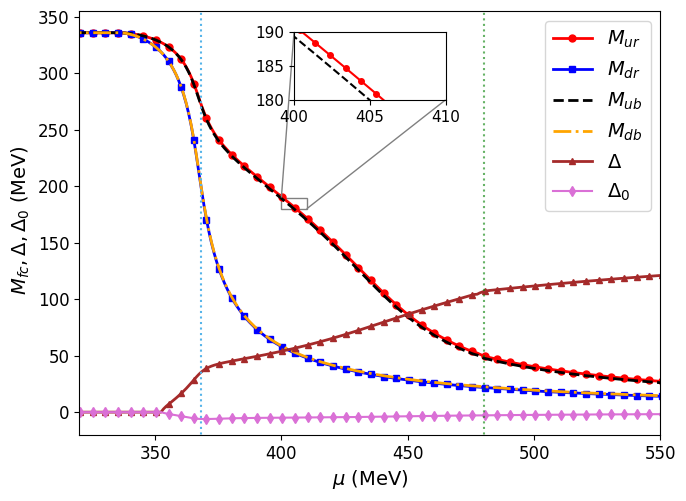}
\caption{The chemical potential dependence of the constituent quark mass $M_{fc}$ and 2SC gap $\Delta,\,\Delta_0$. The $M_{db}$ (orange dash-dotted line) and $M_{dr}$ (blue solid-square line) are coincided. The vertical blue-dotted line at $\mu\approx370 \text{~MeV}$ and green-dotted line at $\mu\approx480\text{~MeV}$ denote, respectively, the locations of the the chiral phase transition and the g2SC to 2SC phase transition. }\label{fig1}.
\end{figure}
\begin{figure}[htpb]
    \centering
\includegraphics[scale=0.55]{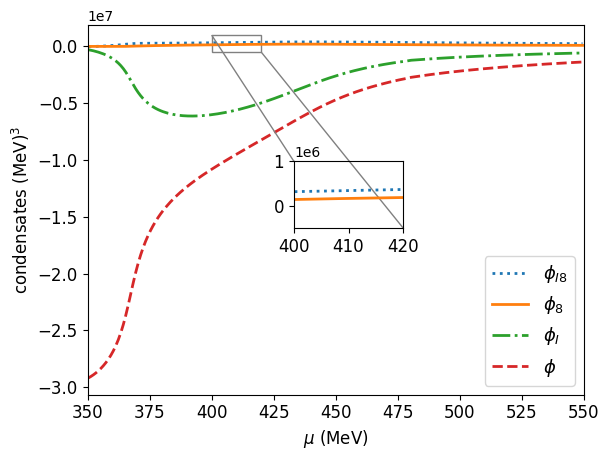}
\caption{The chemical potential dependence of different channel condensates. }
\label{fig2}
\end{figure}

The constituent mass of quarks and 2SC gap are show in Fig.~\ref{fig1}. From this figure we find that, the constituent mass of quarks is nearly independent of color. Explicitly, the mass difference between the red and blue components of up quarks is less than $5$ MeV, while that of down quarks is less than $1$ MeV. These mass differences are related to the condensates $\phi_8$ and $\phi_{I8}$ through $M_{ur}-M_{ub}\sim \phi_8+\phi_{I8}$ and $M_{dr}-M_{db}\sim \phi_8-\phi_{I8}$.  Then, the smallness of mass difference induced by color can be understood from Fig.~\ref{fig2} which shows that both condensates $\phi_8$ and $\phi_{I8}$ are very small. Compared to $\phi_8$ and $\phi_{I8}$, the condensates $\phi_I$ and $\phi$ are big. Thus, we can ignore contributions from the channels $\left(\bar{\psi}\lambda_a \psi\right)^2$ and $\left(\bar{\psi}\tau\lambda_a \psi\right)^2$. The contribution of the $\left(\bar{\psi}\tau \psi\right)^2$ channel, $\phi_I^2$, has a maximum magnitude at about $\mu=390$ MeV, and then decreases with chemical potential, therefore, this channel may have some effects on the 2SC phase transition at relatively low chemical potential (or density), but is expected to vanish as the chemical potential increases. The mass splitting between up and down quarks induced by the $\phi_I$ channel can reach up to 140 MeV, as can be seen from Fig.~\ref{fig1}. In additiotn, this figure also shows that, in the 2SC phase, the 2SC gap parameters $\Delta$ is much larger than $\Delta_0$, agree with that discussed in Ref.~\cite{PhysRevD.65.014018}. 

It is shown in Fig.~\ref{fig1} that, as the chemical potential increases, this model undergoes two transitions. The first is the chiral transition occurring at $\mu \approx 370\text{~MeV}$ which is estimated at the maximum value of the slope of effective mass $M_{fc}$---the susceptibilities~\cite{Du:2013oza}, or equivalently, the slope of $\phi$. In Ref.~\cite{PhysRevD.65.076012}, this transition is first or second order because the chiral condensate is either discontinuous or non-smooth. In our case, however, this transition behaves more like a crossover, as all condensates and densities remain smooth. This difference is due to the vector interaction channel included here, which can make the chiral transition smoother.
The second is the transition from the g2SC phase to the 2SC phase, which occurs at $\mu\approx 480$ MeV according to the criterion~\eqref{eq:Crig2SC}. This result is different from that of Ref.~\cite{PhysRevD.67.065015} which stated that there is a phase transition from 2SC to normal quark phase (NQ). Although the ratio $H/G^{(0)}_s=3/4$ used in Ref.~\cite{PhysRevD.67.065015} is equal to the OGE model considered here, the contribution of channel $(\bar{\psi}\gamma^0\tau \psi)^2$ was not considered. As we will discuss later, this channel will make the 2SC phase more stable than the NQ phase.   

\begin{figure}[htpb]
    \centering
    \includegraphics[scale=0.55]{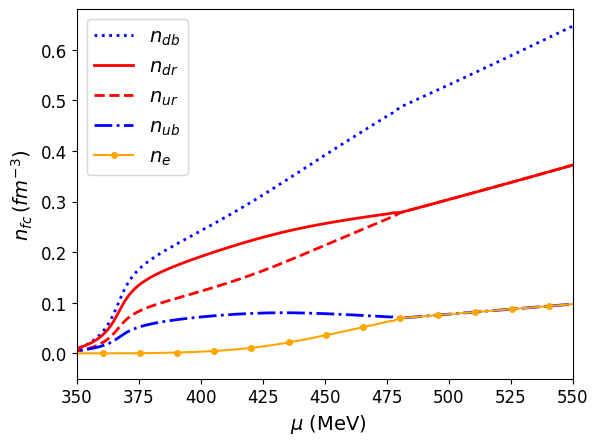}
    \caption{The chemical potential dependence of the quark and electron densities}\label{fig3}
\end{figure}

With respect to the constraints of color and charge neutrality, 
\be
\label{eq19.1}
n_{dr}+n_{ur} & = & n_{ub}+n_{db} ,\nm\\
n_{dr}-n_{ur} & = & n_{ub}-n_e,
\ee
the densities of quarks with different flavors and colors and the density of electrons are displayed in Fig.~\ref{fig3}. One can see that, in the gapless phase, the difference between $n_{ur}$ and $n_{dr}$ first increases then decreases with chemical potential until they are equal where the phase changes from g2SC to gaped 2SC. In the 2SC phase, the mismatch between the fermion surfaces of the $u$-red ($u$-green) quark and $d$-green ($d$-red) quark vanishes, and $n_{ug}=n_{ur} =n_{dg}=n_{dr}$. Thus, $n_{ub}=n_e$. 

\begin{figure}[htpb]
    \centering
    \includegraphics[scale=0.5]{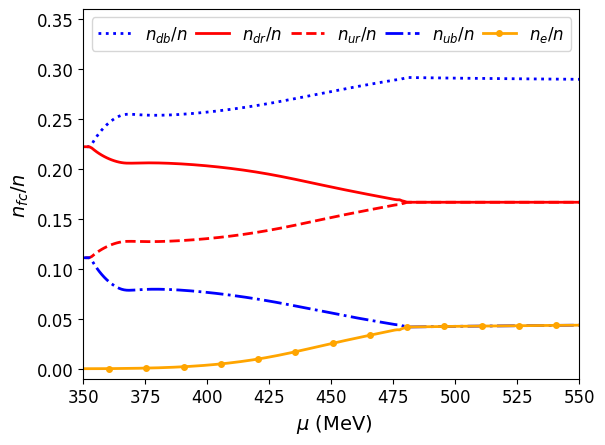}
    \caption{The particle density fractions of quarks and electrons}\label{fig4}
\end{figure}

We plot in Fig.~\ref{fig4} the particle density fractions of quarks and electrons such that the charge neutrality is more clearly illustrated. The charge neutrality requires that 
\begin{equation}
n_{dr}-n_{ub}=n_{db}-n_{ur}=n_{ur}-n_e,
\end{equation}
which is illustrated in Fig.~\ref{fig4} as the gaps between corresponding lines. The interesting thing is that $n_{dr}-n_{ub}$ is roughly proportional to the quark number density $n$ in both the g2SC and 2SC phases. The reason of this observation will be discussed later. 

In the 2SC phase, $n_{ub}=n_{e}$, therefore we have  $\hat{\mu}_{ub}\approx\mu_e=-\mu_Q$ in the chiral limit. Additionally, according the chemical potential relations~\eqref{eq19}, we have 
\begin{equation}
\hat{\mu}_{db}\approx 2\mu_e-4\left(G^{(0)}_{vIv}+\frac{4}{3}G^{(8)}_{vIv}\right)n_I.
\end{equation}
Because of 
\begin{equation}
    n=2(n_{ur}+n_{dr})+n_{ub}+n_{db} =3(n_{ub}+n_{db}),
\end{equation} 
we obtain the fraction of electrons at zero temperature as 
\be
\label{eq23}
\frac{n_e}{n} & = & \frac{1}{3\left[2-4\left(G^{(0)}_{vIv}+\frac{4}{3}G^{(8)}_{vIv}\right)\frac{n_I}{\mu_e}\right]^3+3}\nm\\
& \approx & \frac{1}{27}+\frac{16}{81}\left[G^{(0)}_{vIv}+\frac{4}{3}G^{(8)}_{vIv}\right]\frac{n_I}{\mu_e} \nm\\
& \approx &\frac{1}{27}-\frac{16}{81}\times\frac{7}{3\pi^2}\left[G^{(0)}_{vIv}+\frac{4}{3}G^{(8)}_{vIv}\right]\mu_e^2 \nm\\
& \approx & \frac{1}{27} \nm\\
& &{} -\frac{112}{243\pi^2}\times\left(\frac{3}{5}\right)^2\left[G^{(0)}_{vIv}+\frac{4}{3}G^{(8)}_{vIv}\right]\mu^2.
\ee
If we ignore the second term since the coupling constants $G \sim 1/\Lambda^2$ implies that the second term is much smaller than the first one, we have $n_e/n\approx 1/27$.
The fraction of $n_{db}/n\approx 8/27$ is directly obtained from the neutrality constraints~(\ref{eq19.1}). In 2SC phase, we can obtain the approximate ratios of quarks and electrons as $n_{db}:n_{ur}:n_{dr}:n_{ub}:n_e\approx 8:4.5:4.5:1:1$.  
In the NQ phase, we can also obtain the fractions $n_{dc}/n\approx 2/9$ and $n_{uc}/n\approx 1/9$ for $c=\{ r, g, b\}$. The electrons in NQ phase is negligible, $n_e/n\approx 0$. In the g2SC phase, the particles fractions lie between those of NQ phase and 2SC phase, namely,
\be
        0 & \lesssim & \frac{n_e}{n}\lesssim \frac{1}{27},\nm\\              
        \frac{1}{27} & \lesssim & \frac{n_{ub}}{n}\lesssim \frac{1}{9},\nm\\
        \frac{1}{9} & \lesssim & \frac{n_{ur}}{n}\lesssim \frac{1}{6},\nm\\
        \frac{1}{6} & \lesssim & \frac{n_{dr}}{n}\lesssim \frac{2}{9},\nm\\
        \frac{2}{9} & \lesssim & \frac{n_{db}}{n}\lesssim \frac{8}{27}.
\ee
From these relations, we can obtain that $0.11\lesssim(n_{dr}-n_{ub})/n\lesssim 0.13$ which hardly changes with the chemical potential. In this model, including the second term of Eq.~(\ref{eq23}) results in a slight increase in the electron fraction from $1/27\approx0.037$ to $0.045$ when $\mu=500$~MeV.

\begin{figure}[htpb]
    \centering
    \includegraphics[scale=0.55]{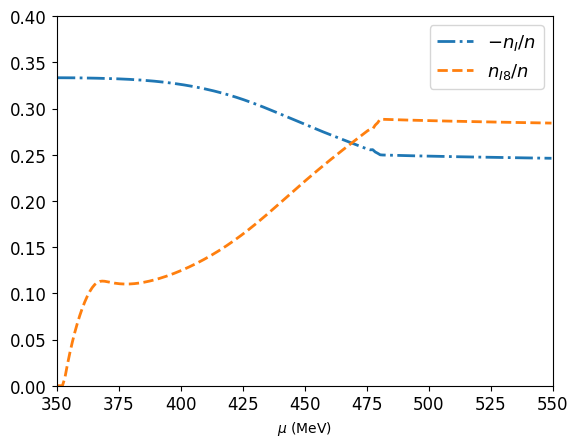}
    \caption{The contributions of $n_I$ and $n_{I8}$ channels.}\label{fig5}
\end{figure}
The contributions of $n_I$ and $n_{I8}$ channels are displayed in Fig.~\ref{fig5}. This figure shows that, in the g2SC phase, $n_I$ is larger than $n_{I8}$, thus the effect of $n_{I8}$ is insignificant, while in the 2SC phase, $n_{I8}$ is proportional to $n_{I}$ as $n_{I8}=-2n_I/\sqrt{3}$ and both effects cannot be neglected. Furthermore, in the gap equations~(\ref{eq19}), the contributions of $n_I$ and $n_{I8}$ are always added together, so the effects of $n_{I8}$ can be mimicked by $n_I$ totally. Considering the above reasons, in the following discussion, we will not address the impact of $n_{I8}$ individually, but focus solely on the effect of $n_{I}$. 

\begin{figure}[htpb]
    \centering
    \includegraphics[scale=0.5]{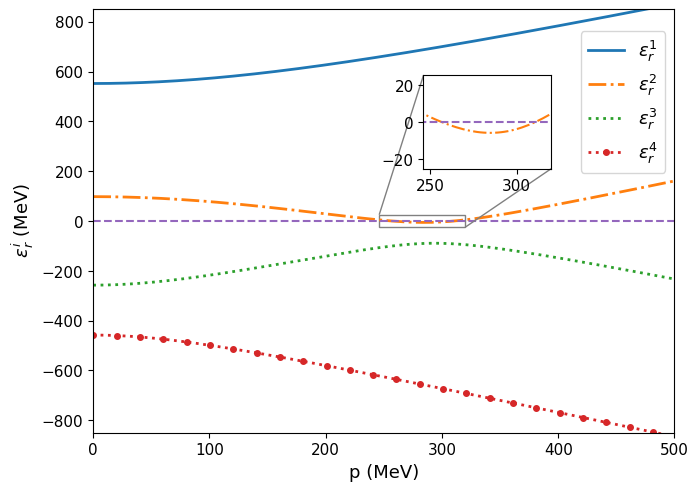}
    \caption{Dispersion relations $\epsilon^i_r(p)$ at $\mu=380~\text{MeV}$.}\label{fig6}
\end{figure}
\begin{figure}[htpb]
\centering
\includegraphics[scale=0.55]{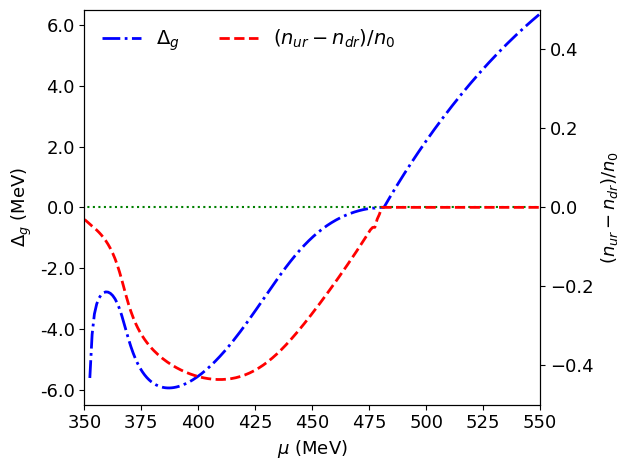}
\caption{ Chemical potential dependence of gap $\Delta_g$ of exciting a quasiparticle and the density difference between the u red and d red quarks. $n_0\simeq 0.16$~fm$^{-3}$ is the nuclear saturation density.}
\label{fig7}
\end{figure}

Fig.~\ref{fig6} displays the spectra $\epsilon^i_r(p)$ of the gap equations at $\mu=380\text{~MeV}$. In this diagram, the $\epsilon^2_r$ and $\epsilon^3_r$ are particle modes, and $\epsilon^1_r$ and $\epsilon^4_r$ are 
antiparticle modes. If we ignore $\Delta_0$ and $\delta M_r$ they are expressed as
\be
\label{eq25}
\epsilon^1_r(p) & = & \sqrt{(E_r(p)+\mu)^2+\Delta^2}-|\delta \mu|,\nm\\
\epsilon^2_r(p) & = & \sqrt{(E_r(p)-\mu)^2+\Delta^2}-|\delta \mu|,\nm\\
\epsilon^3_r(p) & = &{} -\sqrt{(E_r(p)-\mu)^2+\Delta^2}-|\delta \mu|,\nm\\
\epsilon^4_r(p) & = &{} -\sqrt{(E_r(p)+\mu)^2+\Delta^2}-|\delta \mu|.
\ee
Therefore, if there is no mismatch between Fermi surfaces of u-red (u-green) and d-green (d-red), i.e., $\delta \mu=0$, the minimum energy required to break up a Cooper pair and excite a quasiparticle is $\Delta$ (see the $\epsilon^2_r$ in above equations). However, the mismatch of the Fermi surfaces hinders the formation of Cooper pairs and makes it easier to excite a quasiparticle. When $|\delta \mu|\geq \Delta$, i.e., $\epsilon^i_r$ passes through the zero value such as the $\epsilon^2_r$ in Fig.~\ref{fig6}, there is no gap to excite the quasiparticle, this is so-called gapless phase.  We denote the minimal of the $\epsilon^2_r$ in Fig.~\ref{fig6} as $\Delta_g$ and draw its value as a function of $\mu$ in Fig.~\ref{fig7}. From this picture we can see that, in the g2SC phase, the density difference between u red and d red quarks and the gap $\Delta_g$ show the same trend as $\mu$ changes, but former is slightly lagged. 

\section{Influence of different channels on 2SC phase structure}

\label{sec3}

Based on the aforementioned analysis, it becomes evident that, besides the diquark channel $\Delta$ and vector channel $n$, the channels $\phi_I$ and $n_I$ may have considerable influences on 2SC phase diagram. In order to see their distinct impacts on the phase diagram, we systematically modulate the coupling constants associated with these channels. We vary the ratios of the coupling constants of these channels to that of the scalar channel, and separately discuss the effects of these channels while neglecting the contributions from other channels. In this sense, the results in this section is beyond the OGE model.

\subsection{Phase diagram of 2SC}

\begin{figure}[htbp]
    \centering
\includegraphics[scale=0.55]{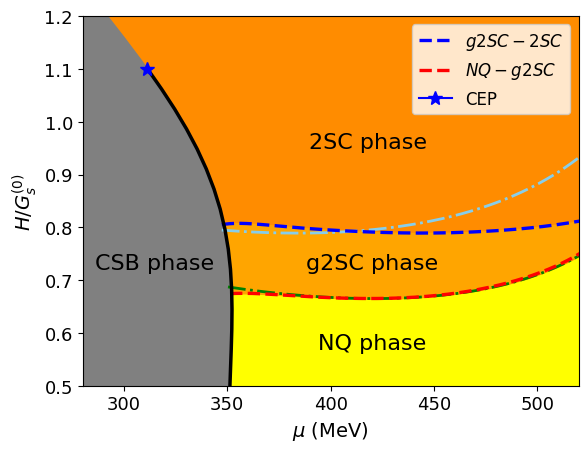}
\caption{The phase diagram in the $\mu$-$H/G_s^{(0)}$ plane. The solid line denotes the first order chiral phase transition, the blue and red dashed lines represent the two boundaries of g2SC phase. For comparison, the boundaries of g2SC phase obtained in Ref. \cite{HUANG2003835} are plotted as dash-dotted lines.}
\label{fig8}
\end{figure}

The phase diagram with different values of $H/G^{(0)}_s$ is shown in Fig.~\ref{fig8}. Here, we only take the diquark channel into consideration. In this figure, we can see that the g2SC phase lies in the range of $0.67\lesssim H/G^{(0)}_s\lesssim0.81$ at low chemical potential $\mu \lesssim 400$~MeV, consistent with Ref.~\cite{HUANG2003835} which was obtained in the chiral limit and with the approximations $\Delta\ll \bar{\mu}\ll \Lambda$. The higher values of $H/G^{(0)}_s$ corresponds to the gaped 2SC phase, while the lower values of $H/G^{(0)}_s$ corresponds to normal quark phase. 
%
This observation indicates that the current quark mass does not affect the lower boundary of the g2SC phase so much. However, the upper bound of the g2SC phase differs somewhat from that of Ref.~\cite{HUANG2003835}. The reason for this discrepancy is that the approximation $\Delta\ll \bar{\mu}\ll \Lambda$ used in Ref.~\cite{HUANG2003835} is no longer valid when $\mu$ approaches the cutoff $\Lambda$.

\begin{figure}[htbp]
\centering
\includegraphics[scale=0.5]{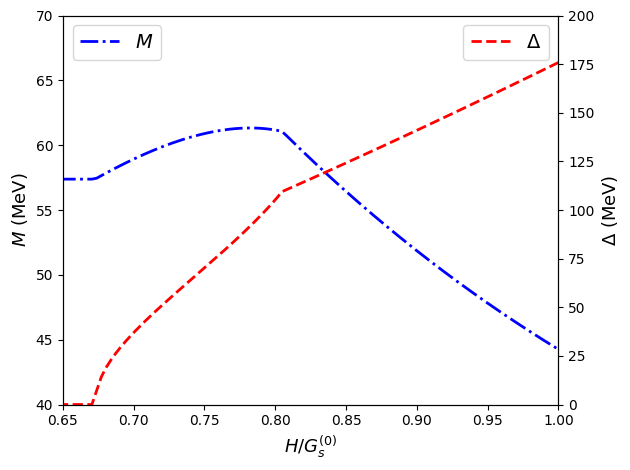}
\caption{$H/G^{(0)}_s$ effect on the constituent quark mass $M$ and diquark gap parameter $\Delta$ at $\mu=370$~MeV. }
\label{fig9}
\end{figure}

\begin{figure}[htbp]
\centering
\includegraphics[scale=0.5]{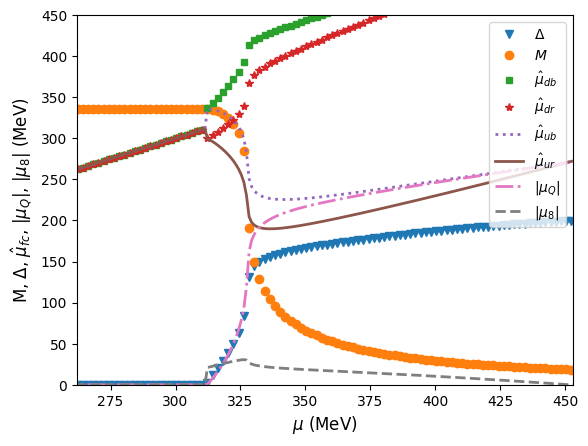}
\includegraphics[scale=0.5]{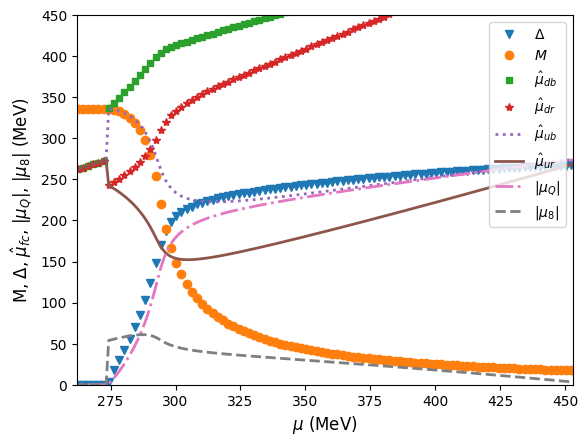}
\caption{The chemical potential dependence of the constituent quark mass $M$, 2SC gap $\Delta$, effective chemical potentials $\hat{\mu}_{fc}$,  $|\mu_Q|$ and $|\mu_8|$ , with different values of $H/G^{(0)}_s$. The upper panel corresponds to $H/G^{(0)}_s = 1.0$, and the lower panel to $H/G^{(0)}_s = 1.2$. }
\label{fig:solutions}
\end{figure}

Moreover, Fig.~\ref{fig8} also shows that the increasing of the value of $H/G^{(0)}_s$ decreases the chemical potential of chiral phase transition.  In order to understand the reason for this effect, we analyze the impact of the diquark channel with fixed chemical potential in Fig.~\ref{fig9}. This figure illustrates that in the g2SC phase, the diquark gap parameter $\Delta$ slightly increases the constituent quark mass, while in the 2SC phase, it decreases the constituent quark mass. For this reason, in 2SC phase, the increasing of $\Delta$ will cause the phase transition to occur earlier. Especially, when $H/G^{(0)}_s$ is larger than $1.1$, the chiral phase transition is crossover instead of first-order. This result is quite different from Ref.~\cite{PhysRevD.65.076012}, where the charge and color neutrality constraints were not taken into account. As shown in Fig.~\ref{fig:solutions}, the chiral phase transition is of first order when \( H/G^{(0)}_s = 1.0 \). As \( H/G^{(0)}_s \) increases, the transition becomes a crossover, rather than a second order phase transition suggested in Ref.~\cite{PhysRevD.65.076012}. In our case---for example, at \( H/G^{(0)}_s = 1.2 \)---all quantities appear to vary smoothly with the chemical potential in the vicinity of the chiral phase transition, indicating that the transition is a crossover.
The CEP is located at $\mu=312{\rm~MeV}, H/G^{(0)}_s=1.1$, and the corresponding $\Delta =138\,\text{MeV}$.

\subsection{Effects of other channels on 2SC phase structure}

In order to show the influences of the different channels, explicitly, the $\phi_I$, $n$, and $n_I$ channels, on the phase diagram, we separately solve the gap equations with some typical values of couplings.

\begin{figure}[htpb]
    \centering
    \includegraphics[scale=0.55]{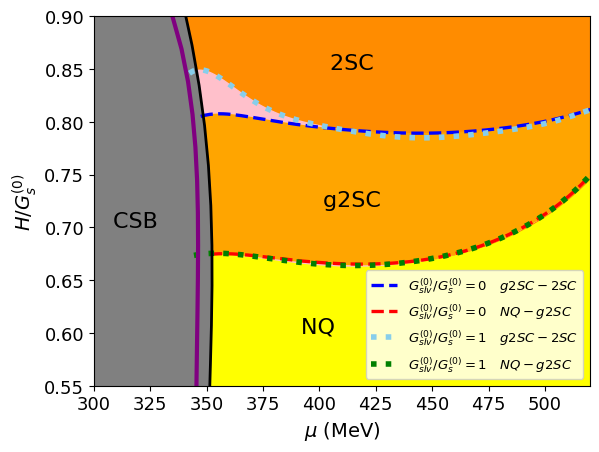}
    \caption{Phase diagram in the $\mu$-$H/G_s^{(0)}$ plane. The black and purple solid lines correspond to the first order chiral phase transitions with and without considering the $\phi_I$ channel interactions, respectively. The dashed and dotted lines represent the phase boundaries of the g2SC phase, with and without the $\phi_I$ channel contributions, respectively.}\label{fig10}
\end{figure}

In Fig.~\ref{fig10} we analyze the effect of $\phi_I$ channel with $G^{(0)}_{sIv}/G^{(0)}_s=1$. We find that the $\phi_I$ channel decreases the chemical potential of chiral phase transition slightly. In addition, it expands the gapless phase's range from $0.67-0.8$ to $0.67-0.85$ in the chemical potential region shortly after the chiral phase transition. The $\phi_I$ channel mainly affects the mass difference between up and down quarks. This mass difference contributes a value of $M_r\delta M_r/E_r(p)$ to $|\delta\mu|$  in dispersion relation $\epsilon^2_r(p)$ in Eq. (\ref{eq25}), transforming the original gaped 2SC into the g2SC region. Meanwhile, the $\phi_I$ channel hardly changes the lower boundary of the g2SC phase and the upper boundary at high chemical potential.

\begin{figure}[htpb]
    \centering
    \includegraphics[scale=0.55]{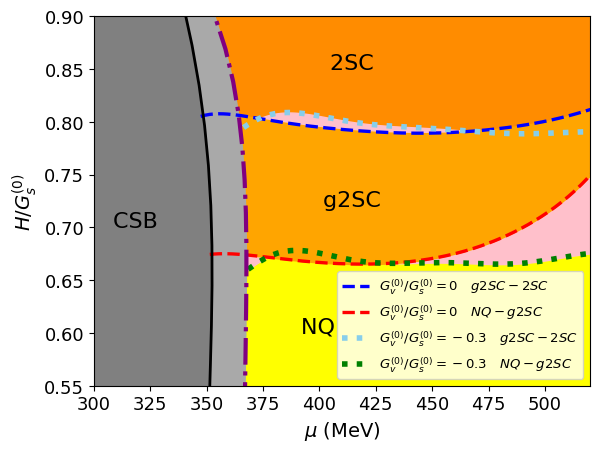}
    \caption{Effects of $n$ channel on the phase boundaries of the g2SC phase. The purple dash-dotted line denotes the crossover chiral phase transition. The dotted and dashed lines indicate the g2SC phase boundaries with and without the $n$ channel interactions, respectively. }\label{fig11}
\end{figure}

The effect of $n$ channel on 2SC phase are shown in Fig.~\ref{fig11} with $G^{(0)}_v/G^{(0)}_s={}-0.3$. We choose a negative $G^{(0)}_v$ because, after bosonizing the NJL Lagrangian, ${}-1/G^{(0)}_v$ is proportional to the squared mass of the vector meson times a positive constant~\cite{Braun:2011pp}. The main feature of the $n$ channel is to shift the entire phase diagram along the high chemical potential direction. At the same time, the type of chiral phase transition changes from a first-order phase transition to a crossover. From the gap equation~\eqref{eq19} we can see that, due to the negative coupling $G^{(0)}_v$, the density $n$ decreases the effective chemical potential, which is equivalent to an increase of the chemical potential. Consequently, this channel shifts the entire diagram to the right. As mentioned in Ref.~\cite{PhysRevD.67.065015}, when the chemical potential nears the cutoff $\Lambda$, the diquark condensate starts to be affected by the momentum cutoff, declining to zero, and eventually triggers a transition to the NQ phase. The $n$ channel's contribution removes this limitation, stabilizing the 2SC phase at high chemical potentials. 

\begin{figure}[htpb]
    \centering
    \includegraphics[scale=0.55]{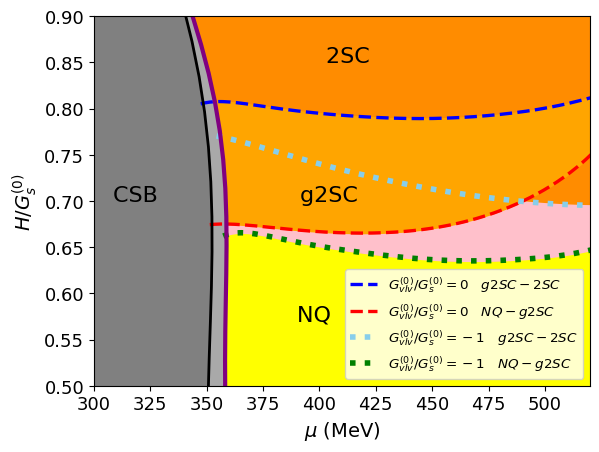}
    \caption{Comparison of the g2SC phase boundaries with different $n_I$ channel coupling strengths. The dashed and dotted lines correspond to the g2SC phase boundaries without and with the inclusion of the $n_I$ channel contributions, respectively}\label{fig12}
\end{figure}

Finally, we show the effects of the $n_I$ channel in Fig.~\ref{fig12} by using $G^{(0)}_{vIv}/G^{(0)}_s=-1$. This figure shows that the $n_I$ channel reduces the upper and lower boundaries of the g2SC phase, with the effect more significant as the chemical potential increases. The gap equation~(\ref{eq19}) implies that the mismatch $\delta\mu$ of the chemical potential between the u-red and d-green quarks is proportional to $G^{(0)}_{vIv}n_I$. Because we take $G^{(0)}_{vIv}<0$, $n_I$ decreases the value of $|\delta\mu|$, then decreases the g2SC upper boundary. As we mentioned above, there is a phase transition from g2SC to 2SC instead of to NQ phase at $H/G^{(0)}_s\sim 0.75$. The g2SC to NQ phase transition disappears, replaced by a g2SC to 2SC phase transition.

\section{Summary and discussion}
\label{sec4}

In this paper, we systematically investigated the effects of various interaction channels within the NJL model on the 2SC phase at zero temperature by seriously considering the constraints from color neutrality and charge neutrality. We started from the general NJL model and derived the relations among the couplings by using Fierz transformations. All these couplings are controlled by only five independent couplings. 

To figure out the dominant channels, we adopted the traditional OGE model. The results showed that in addition to the diquark and vector channels widely discussed in the literature, due to the inclusion of both charge/color neutrality constraints and self-consistent interaction contributions, scalar-isovector, vector-isovector, and vector-isovector-color-octet channels are also found important, while other channels can be neglected. Furthermore, our analysis showed that the color and charge neutrality conditions give rise to the emergency of the g2SC phase between the NQ and 2SC phases, while the contribution from the $\phi_I$ channel induces the phase transition from g2SC to 2SC. The particle fractions of quarks with different colors and flavors were estimated as $n_{db}:n_{ur}:n_{dr}:n_e\approx 8:4.5:4.5:1:1$ in the 2SC phase. Moreover, the $\phi_{sIv}$ channel results in a mass splitting of up to 140 MeV between the $u$ and $d$ quarks. Additionally, the contribution from the $n$ channel turns the chiral phase transition into a crossover, rather than a first-order transition.

Next, we discussed the impacts of some channels on the phase structure beyond the OGE approximation. We found that the gapless phase lies in the range of $0.67\lesssim H/G^{(0)}_s\lesssim 0.81$ which agrees with the results of Ref.~\cite{HUANG2003835}. Moreover, the increasing value of $H/G^{(0)}_s$ decreases the chemical potential of chiral phase transition. This result is caused by the competition between the diquark condensate and chiral condensate in the 2SC phase~\cite{PhysRevD.65.076012,PhysRevD.108.043008}. In addition, we found the CEP on the $\mu$--$H/G^{(0)}_s$ plane located at $\mu=312$ MeV, $H/G^{(0)}_s=1.1$ with $\Delta=138$ MeV. The chiral transition is crossover instead of first order when $H/G^{(0)}_s>1.1$. For the influences of other channels on 2SC phase diagram, we found that the scalar-isovector channel $\phi_I$ expand the gapless phase from $0.67\lesssim H/G^{(0)}_s\lesssim 0.8$ to $0.67\lesssim H/G^{(0)}_s\lesssim 0.85$ in the chemical potential region shortly after the chiral phase transition. The vector channel shifts the entire phase diagram along the larger $\mu$ direction. The isovector channel reduces the values of $H/G^{(0)}_s$ of both the upper and lower boundaries of the g2SC phase.

The above results indicate that when studying color-superconducting phases and their applications in neutron star physics, these interaction channels should also be considered. The effects of these channels on the finite-temperature color-superconducting phase diagram and the equation of state for neutron stars will be investigated in the future work.

\section*{Acknowledgements} 

The work of Y.~L. M. is supported in part by the National Science Foundation of China (NSFC) under Grant No. 12347103 and Gusu Talent Innovation Program under Grant No. ZXL2024363. The work of C.~M. L. is supported in part by the National Science Foundation of China (NSFC) under Grant No. 12005192 and the Natural Science Foundation
of Henan Province of China (No. 242300421375).

\bibliography{refs}

\end{document}